\documentclass[12pt]{article}
\def\bea{\begin{eqnarray}}
\def\ve{\vert}
\def\eea{\end{eqnarray}}

\def\nnb{\nonumber}
\def\la{\langle}
\def\ra{\rangle}
\def\ba{\begin{array}}
\def\ea{\end{array}}
\def\tr{\mbox{Tr}}

\def\xis0{{\Xi^{*0}}}

\def\ss{\la \bar s s \ra}

\def\qq{\la \bar q q \ra}
\def\gGgG{\la g^2 G^2 \ra}
\def\q{\gamma_5 \not\!q}
\def\x{\gamma_5 \not\!x}
\def\sig{\sigma_{\mu \nu} \gamma_5 p^\mu q^\nu}

\begin{document}
\title{\bf Meson-Baryon Couplings and the F/D ratio in Light Cone QCD}
\author{T.M. Aliev, A. {\"O}zpineci, M. Savc{\i} \\
Physics Dept. Middle East Technical University \\
06531 Ankara,Turkey}
\begin{titlepage}
\maketitle
\begin{abstract}

Using the general form of the baryon currents, we calculate the meson-baryon coupling constants
and the F/D ratio within the framework of light cone QCD sum rules in the $SU(3)$ flavor symmetry
limit. The dependence of the results on the Dirac structure and on the free parameter $b$ appearing in
the general baryon current is considered. Comparison of our results on F/D ratio with the existing
results is presented.
\end{abstract} 
\end{titlepage}

\section{Introduction}
Determination of the various fundamental parameters of hadron from experimental data requires
information about physics at large distance.  Unfortunately such information can not be achieved from
the first principles of a fundamental theory of strong interactions due to its very complicated infrared
behavior.  For this reason to determine properties of hadrons a reliable non-perturbative approach is
needed.  Among all non-perturbative approaches, QCD sum rules \cite{shifman} are especially powerful in
investigating the properties of low lying hadrons.  Note that in traditional sum rules non-perturbative
effects are taken into account through various condensates.  Among various applications, determination
of meson baryon couplings is of particular interest, since they are main ingredients of baryon-baryon
interactions.  Recently, the two point correlation function of the nucleon with a pion
\bea
\Pi = i \int d^4 x e^{i p x} \la \pi(q) \ve {\cal T} \eta_N(x) \bar \eta_N (0) \ve 0 \ra
\eea
has been extensively used to calculate pion-nucleon coupling in the framework of QCD sum
rules [2-9]. 

This correlator function contains three different Dirac structures: a) $i \gamma_5$; b) $\gamma_5
\sigma_{\mu \nu} q^\mu p^\nu$; and c) $i \gamma_5 \not\!p$, each of which can in principle be used to
calculate the meson-baryon coupling.  In \cite{kim}, it is shown that the predicted pion-nucleon coupling
depend on the Dirac structure.  In \cite{kim2, kim3} sum rules for the $i \gamma_5$ structure and in \cite{kim4} for
$\gamma_5 \sigma_{\mu \nu} q^\mu p^\nu$    structure beyond the chiral limit were obtained.  Both sum
rules yield the $\pi NN$ coupling close to its experimental value, while the $i \gamma_5  \not\!p$ sum
rules contain large contributions from the continuum and for this reason its predictions are not reliable.  
The pseudoscalar and tensor sum rules beyond the chiral limit have been applied to other
meson-baryon couplings $\eta NN$, $\pi \Xi \Xi$, $\pi \Sigma \Sigma$, and $\eta \Sigma \Sigma$
\cite{kim3, kim4}.  It is well known that the meson-baryon couplings in $SU(3)$
limit can be classified in terms of two parameters: the $\pi NN$ coupling and the so called F/D ratio
\cite{swart}.  The sum rules for the F/D ratio for general form of baryon currents for above mentioned
three different Dirac structures in the framework of traditional sum rules is studied in \cite{doi}.

In this work, our aim is to study F/D ratio in the framework of an alternative approach to the
traditional sum rules, namely light cone QCD sum rules(LCQSR) and compare the predictions of these
different approaches.  LCQSR is based on the operator product expansion on the light cone, which is an
expansion over the twists of the operators rather than the dimensions in the traditional sum rules and
the main contribution comes from the lowest twist operators.  The matrix elements of the nonlocal
operators sandwiched between a hadronic state and the vacuum defines the hadronic wave functions (about
the light cone QCD sum rules see \cite{braun, braun2} and references therein).

The paper is organized as follows.  In Sect. II, the light cone sum rules for meson baryon coupling
using general baryon currents for the structures $i \gamma_5 \not\!q$ and $\sig$  are obtained.  More
over, we construct an expression for the F/D
ratio from OPE. Sect. III contains our numerical results and conclusions.

\section{Sum Rules for the Meson-Baryon Couplings and the F/D Ratio}
As we have noted, meson-baryon couplings in the $SU(3)$ flavor symmetry limit can be expressed in terms
of only two parameters \cite{swart} $g_{\pi NN}$ and $\alpha=\frac{F}{F+D}$ as:
\bea
g_{\eta NN} = \frac{1}{\sqrt{3}} (4 \alpha -1) g_{\pi N N} \, , &;&
g_{\pi \Xi \Xi} = (2 \alpha-1) g_{\pi N N} \, , \nnb \\
g_{\eta \Xi \Xi} = -\frac{1}{\sqrt{3}} (1 + 2 \alpha) g_{\pi N N} \, , &;& 
g_{\pi \Sigma \Sigma} = 2 \alpha g_{\pi N N} \, , \nnb \\
g_{\eta \Sigma \Sigma} = \frac{2}{\sqrt{3}} (1-\alpha) g_{\pi N N} \, , && \label{su3}
\eea

In this section, we will derive light cone sum rules for the $\pi^0 N N$ coupling.
A sum rule for the meson-baryon couplings can be constructed by equating two different representations
of a suitably chosen correlator, written in terms of hadrons and quark-gluons. 
We  begin our calculations by considering the following correlator:
\bea
\Pi = i \int dx e^{i p x} \la {\cal M}(q) \ve {\cal T} J^B(x) \bar J^B(0) \ve 0
\ra \label{correlation}\, ,
\eea
where $J^B$ is the current of the baryon under consideration, ${\cal T}$ is the time ordering operator,
$q$ is the momentum of the meson ${\cal M}$. This correlator can be calculated on one side
phenomenologically, in terms of the hadron parameters, and on the other side by the operator product
expansion (OPE) in the deep Euclidean region, $p^2 \rightarrow - \infty$, using QCD degrees of freedom.
By equating both expressions, we construct the corresponding sum rules.

Saturating the correlator, Eq. (\ref{correlation}), by ground state baryons we get:
\bea
\Pi (p_1^2,p_2^2) = \frac{\la 0 \ve J^B \ve B_1(p_1) \ra}{p_1^2 - M_1^2} 
\la B_1 (p_1) {\cal M}(p) \ve  B_2(p_2) \ra 
\frac{\la B_2 (p_2) \ve J^B \ve 0 \ra}{p_2^2 - M_2^2},
\label{insert}
\eea 
where $p_2 = p_1 + q$ and $M_i$ is the mass of the baryon $B_i$.


The matrix elements of the interpolating currents between the ground state and
the state containing a single baryon, $B$, with momentum $p$ and having spin $s$ is defined as:
\bea
\la 0 \ve J^B \ve B(p,s) \ra = \lambda_B u(p,s) \, , \label{3}
\eea
where $\lambda_B$ is the residue, and $u$ is the Dirac spinor for the baryon.
In order to write down the phenomenological part of the sum rules from Eq. (\ref{insert}) it follows
that one also needs an expression for the matrix element $\la B(p_1) {\cal M}\ve B (p_2)$. 
This matrix element is defined as:
\bea
\la B(p_1) {\cal M}\ve B (p_2) \ra =   g_{{\cal M}B B}\bar u(p_1) i \gamma_5 u(p_2)
\eea
With these definitions, the phenomenological representation of the correlator becomes:
\bea
\Pi_{{\cal M} BB }(p_1,p_2) = - \frac{\lambda_B^2 g_{{\cal M} B B}}{(p_1^2
- m_B^2)(p_2^2 - m_B^2)} \left( 
p_1^\mu q^\nu \sigma_{\mu \nu} \gamma_5 - m_B i \gamma_5 \not\!q + \frac{m_\pi^2}{2} i \gamma_5 \right) 
+ \ldots
\label{phenomenological}
\eea 
where $\ldots$ stands for the contribution of higher states and the continuum.  Note that in this work, 
we will consider massless pion in which case, the pseudo scalar structure $i \gamma_5$ vanishes
identically and hence we will omit this structure.  In this work we will both of the remaining 
structures appearing in Eq. (\ref{phenomenological}) and compare the reliability of the structures.

On the QCD side, in order to evaluate the correlator, one needs suitable expressions for the baryon
currents.  In this work, we will use the following general forms of baryon currents:
\bea
J_p(x,t) &=& 2 \epsilon_{abc} \left\{ \left[ u_a^T(x) C d_b(x) \right] \gamma_5 u_c(x) + 
b  \left[ u_a^T(x) C \gamma_5 d_b(x) \right]  u_c(x) \right\} \, , \\
J_\Xi(x,t) &=&- 2 \epsilon_{abc} \left\{ \left[ s_a^T(x) C u_b(x) \right] \gamma_5 s_c(x) + 
b  \left[ s_a^T(x) C \gamma_5 u_b(x) \right]  s_c(x) \right\} \, ,  \\
J_\Sigma(x,t) &=& 2 \epsilon_{abc} \left\{ \left[ u_a^T(x) C s_b(x) \right] \gamma_5 u_c(x) + 
b  \left[ u_a^T(x) C \gamma_5 s_b(x) \right]  u_c(x) \right\} \,
\eea
where $C$ is the charge conjugation operator and $b$ is an arbitrary real parameter.  The Ioffe current
corresponds to the choice $b=-1$. The currents for $\Xi$ and $\Sigma$ can be obtained from the nucleon
interpolating current via the $SU(3)$ rotation.  Namely $\Xi$ and $\Sigma$ currents can be easily
obtained from the nucleon current by the following simple replacements: to obtain the $\Xi$ current,
substitute $s$ and $u$ in place of $u$ and $d$ respectively and to obtain the $\Sigma$ current,
substitute $s$ in place of $d$ in the proton current. Note that the pion-nucleon coupling constant for
the Ioffe current in LCQSR have been calculated in \cite{aliev}.

In the large Euclidean momentum $-p^2 \rightarrow \infty$ region, the correlator can be
calculated using the OPE. For the pion-proton sum rule, the OPE yields:
\bea
\Pi_{\pi N N}(p_1,p_2) &=&
i \epsilon_{a b c} \epsilon_{d e f} \int d^4x e^{i p x} \la \pi \ve
\bar u^d A_i u^a \left\{
\gamma_5 A_i \gamma_5 \tr S_d^{be} {S'}_u^{cf} + 
\right. \nnb \\ 
&+&\left. \gamma_5 A_i {S'}_d^{be} S_u^{cf} \gamma_5 +
b^2 A_i \gamma_5 {S'}_d^{be} \gamma_5 S_u^{cf} +
\right. \nnb \\ 
&+& \left. b^2 A_i \tr S_u^{cf} \gamma_5 {S'}_d^{be} \gamma_5 +
b \gamma_5 A_i \gamma_5 {S'}_d^{be} S_u^{cf} + 
\right. \nnb \\ 
&+& \left. b \gamma_5 A_i \tr S_u^{cf} \gamma_5 {S'}_d^{be} + 
b A_i {S'}_d^{be} \gamma_5 S_u^{cf} \gamma_5 +
\right. \nnb \\ 
&+&\left. b A_i \gamma_5 \tr S_u^{cf} {S'}_d^{be} \gamma_5 +
\gamma_5 S_u^{cf} \gamma_5 \tr S_d^{be} A_i' + 
\right. \nnb \\ 
&+&\left. \gamma_5 S_u^{cf} {S'}_d^{be} A_i \gamma_5 +
b^2 S_u^{cf} \gamma_5 {S'}_d^{be} \gamma_5 A_i +
\right. \nnb \\ 
&+&\left. b^2 S_u^{cf} \tr A_i \gamma_5 {S'}_d^{be} \gamma_5 +
b \gamma_5 S_u^{cf} \gamma_5 {S'}_d^{be} A_i + 
\right. \nnb \\ 
&+& \left. b \gamma_5 S_u^{cf} \tr A_i \gamma_5 {S'}_d^{be} + 
b S_u^{cf} {S'}_d^{be} \gamma_5 A_i \gamma_5 +
\right. \nnb \\ 
&+&\left. b S_u^{cf} \gamma_5 \tr A_i {S'}_d^{be} \gamma_5 
\right\} + \nnb \\ 
&+&\bar d^e A_i d^b \left\{
\gamma_5 S_u^{cf} \gamma_5 \tr A_i {S'}_u^{ad} + 
\right. \nnb \\ 
&+&\left. \gamma_5 S_u^{cf} A_i' S_u^{ad} \gamma_5 +
b^2 S_u^{cf} \gamma_5 A_i' \gamma_5 S_u^{ad} +
\right. \nnb \\ 
&+& \left. b^2 S_u^{cf} \tr S_u^{ad} \gamma_5 A_i' \gamma_5 +
b \gamma_5 S_u^{cf} \gamma_5 A_i' S_u^{ad} + 
\right. \nnb \\ 
&+&\left. b \gamma_5 S_u^{cf} \tr S_u^{ad} \gamma_5 A'_i + 
b S_u^{cf} A_i' \gamma_5 S_u^{ad} \gamma_5 +
\right. \nnb \\ 
&+&\left. b S_u^{cf} \gamma_5 \tr S_u^{ad} A_i' \gamma_5
\right\}  \ve 0 \ra \label{proton.pion}
\eea
where $A_i = 1, \, \gamma_\alpha, \, \sigma_{\alpha \beta}/\sqrt{2}, \, i \gamma_\alpha \gamma_5,
\, \gamma_5$, a sum over $A_i$ implied, $S' \equiv CS^TC$, $A'_i = CA_i^TC$, with $T$ denoting the
transpose of the matrix, and $S_q$ is the full quark propagator with both perturbative and
non-perturbative contributions.  In our calculations, we will neglect the masses of the quarks and
assume an $SU(3)$ flavor symmetry.  From Eq. (\ref{proton.pion}) it follows that in order to calculate
the correlator (\ref{proton.pion}), the explicit expression of the massless quark propagator is needed.
The complete light cone expansion of the light quark propagator $S_q$ in external field is 
given in \cite{balitsky}. It gets contributions from the $\bar q G q$, $\bar q GG q$, $\bar q q \bar q q$ 
non-local operators (where $G$ is the gluon field strength tensor).  In this work we consider only operators with one
gluon field, corresponding to the quark-antiquark-gluon components of the pion and neglect components with two gluons or four 
quark fields.  This is consistent with the approximation of the twist 4 two particle wave functions obtained in 
\cite{braun05}. Taking into  account higher Fock-state components would demand corresponding modifications
in the two particle wavefunctions via the equations of motion.  Formally neglect of the $\bar q GG q$, 
$\bar q q \bar q q$ terms can be justified on the basis of an expansion in conformal spin \cite{braun05}.
In this approximation the massless quark propagator is given by:
\bea
S_q &=& \la 0 \ve {\cal T} \bar q(0) q(x) \ve 0 \ra \nnb \\
&=& \frac{i \not\!x}{2 \pi^2 x^4}  -
\frac{\la \bar q q \ra}{12}
 - \frac{x^2}{192}m_0^2 \la \bar q q \ra  - \nnb \\
&-& i g_s \int_0^1 dv \left[ \frac{\not\!x}{16 \pi^2 x^2} G_{\mu \nu}(v x)  \sigma_{\mu \nu} - v x_\mu
G_{\mu \nu}(v x) \gamma_\nu \frac{i}{4 \pi^2 x^2} \right] + \ldots \label{12}
\eea
Note that the local part of the propagator consisting of operators with dimension $d >  5$ is neglected
since they give  a negligible contribution. 
In order to evaluate Eq. (\ref{proton.pion}) analytically, one needs the matrix elements of nonlocal
operators between the pion state and the vacuum state.  The non-zero matrix elements are defined in
terms of the pion wave functions up to twist 4 as (see \cite{braun0, braun05, belyaev}):   
\bea
&&\la \pi(q) \ve \bar u(x) \gamma_\mu \gamma_5 u(0) \ve 0 \ra = \nnb \\ 
&&-i f_\pi q_\mu \int_0^1 du e^{i u q x} \left[ \varphi_\pi(u) + x^2 g_1(u) \right] + 
f_\pi \left(x_\mu - \frac{x^2 q_\mu}{qx}\right) \int_0^1 du e^{i u q x} g_2(u) \, ,
\label{matrix1} \nnb \\
\\
&&\la \pi(q) \ve \bar u(x) i \gamma_5 u(0) \ve 0 \ra = \frac{f_\pi m_\pi^2}{2 m_q} 
\int_0^1 du e^{i u q x} \varphi_P (u) \, , \\
&&\la \pi(q) \ve \bar u(x) \sigma_{\mu \nu} \gamma_5 u(0) \ve 0 \ra = i (q_\mu x_\nu - q_\nu x_\mu) \frac{f_\pi m_\pi^2}{12 m_q} 
\int_0^1 du e^{i u q x} \varphi_\sigma (u) \, , \\
&&\la \pi(q) \ve \bar u(x) \sigma_{\alpha \beta} \gamma_5 g_s G_{\mu \nu}(u x) u(0) \ve 0 \ra =
\nnb \\
&&i f_{3 \pi} \left[ \left(q_\mu q_\alpha g_{\nu \beta} - q_\nu q_\alpha g_{\mu
\beta}
\right) - \left(q_\mu q_\beta g_{\nu \alpha}- q_\nu q_\beta g_{\mu \alpha} \right) \right] \times  
\nnb \\ 
&&\int {\cal D} \alpha_i \varphi_{3 \pi} (\alpha_i) e^{i q x(\alpha_1 + u \alpha_3)} \, , 
\\
&&\la \pi(q) \ve \bar u(x) \gamma_\mu \gamma_5 g_s G_{\mu \nu}(u x) u(0) \ve 0 \ra =
\nnb \\
&&f_\pi \left[ q_\beta \left( g_{\alpha \mu} - \frac{x_\alpha q_\mu}{qx} \right) -
q_\alpha \left( g_{\beta \mu} - \frac{x_\beta q_\mu}{qx} \right) \right] \int {\cal D} \alpha_i
\varphi_\perp(\alpha_i) e^{i q x (\alpha_1 + u \alpha_3)} + \nnb \\
&&+ f_\pi \frac{q_\mu}{qx} \left( q_\alpha x_\beta - q_\beta x_\alpha \right) \int
{\cal D}
\alpha_i \varphi_\parallel(\alpha_i) e^{i q x (\alpha_1 + u \alpha_3)} \, , \\
&&\la \pi(q) \ve \bar u(x) \gamma_\mu g_s {\tilde G}_{\mu \nu}(u x) u(0) \ve 0 \ra  =
\nnb \\
&& i f_\pi \left[ q_\beta \left( g_{\alpha \mu} - \frac{x_\alpha q_\mu}{qx} \right) -
q_\alpha \left( g_{\beta \mu} - \frac{x_\beta q_\mu}{qx} \right) \right] \int {\cal D} \alpha_i
\tilde\varphi_\perp(\alpha_i) e^{i q x (\alpha_1 + u \alpha_3)} + \nnb \\
&&+ i f_\pi \frac{q_\mu}{qx} \left( q_\alpha x_\beta - q_\beta x_\alpha\right) \int {\cal
D}
\alpha_i \tilde\varphi_\parallel(\alpha_i) e^{i q x (\alpha_1 + u \alpha_3)} \, \label{matrix5}.
\eea
Here, the operator $\tilde G_{\alpha \beta}$ is the dual of the gluon field strength tensor, $\tilde
G_{\alpha \beta} = \frac{1}{2} \epsilon_{\alpha \beta \delta \rho} G^{\delta \rho}$, ${\cal
D}\alpha_i$ is defined as ${\cal D}\alpha_i = d\alpha_1 d\alpha_2 d\alpha_3
\delta(1-\alpha_1-\alpha_2-\alpha_3)$.

Note that the corresponding matrix elements for the $d$-quark can be obtained from the $u$-quark matrix
elements using the isospin relation:
\bea
\la \pi \ve \bar d(x) {\cal O} d (0) \ve 0 \ra = - \la \pi \ve \bar u(x) {\cal O} u(0) \ve 0 \ra \, ,
\eea
where ${\cal O}$ is any of the matrices appearing in Eqs. (\ref{matrix1}-\ref{matrix5}). 

Note that one can write Eq. (\ref{proton.pion}) as the sum of two terms:
\bea
\Pi_{\pi NN} = \Pi_1 + \Pi_2
\eea
where $\Pi_1$ is obtained from the terms in the OPE proportional to $\bar u A_i u$ and $\Pi_2$ is
obtained from the terms proportional to $\bar d A_i d$. With this  separation of the terms, and noting
that the currents for the other baryons, $\Sigma$ and $\Xi$ can be obtained from proton current by
simple substitutions, one can immediately obtain the following results for the OPE for the $\pi \Sigma
\Sigma $ and $\pi \Xi
\Xi$ couplings:
\bea
\Pi_{\pi \Sigma \Sigma} &=& \Pi_1 \label{20} \, ,\\
\Pi_{\pi \Xi \Xi} &=& - \Pi_2 \label{21} \, .
\eea
In deriving Eqs. (\ref{20}) and (\ref{21}), we have used the fact that up to twist four the matrix
elements of the $s$-quark operators between
vacuum state and the one pion state are zero.

Extension of the sum rules for the couplings $\pi NN$ to the $\eta NN$ coupling is obvious, i.e. it is
enough in the correlator (\ref{correlation}) to replace $\pi^0$ by $\eta$.  Note that in the $SU(3)$
limit,
there is no $\eta-\eta'$ mixing, and the strange quark component of $\eta$ does not participate in the
sum rules for $\eta NN$, but it gives contribution to the $\eta \Xi \Xi$ and $\eta \Sigma \Sigma$ sum
rules. In this limit,
the matrix elements for $\eta$ can be obtained from the corresponding matrix elements for $\pi$
by:
\bea
\la \eta \ve  \bar u {\cal O} u \ve 0 \ra = 
\la \eta \ve  \bar d {\cal O} d \ve 0 \ra =
\frac{1}{\sqrt{3}} \la \pi \ve  \bar u {\cal O} u \ve 0 \ra \\
\la \eta \ve  \bar s {\cal O} s \ve 0 \ra =
-\frac{2}{\sqrt{3}} \la \pi \ve  \bar u {\cal O} u \ve 0 \ra
\eea
Compared with the $\pi^0$ case, we see that the sign of the contribution coming from the $u$ and $d$
quarks are the same.

The corresponding OPEs can be written as:
\bea
\Pi_{\eta N N} &=& \frac{1}{\sqrt{3}} \left( \Pi_1 - \Pi_2 \right) \, , \\
\Pi_{\eta \Sigma \Sigma} &=& \frac{1}{\sqrt{3}} \left( \Pi_1 + 2 \Pi_2 \right) \, , \\
\Pi_{\eta \Xi \Xi} &=& \frac{1}{\sqrt{3}} \left( -2 \Pi_1 - \Pi_2 \right) \, .
\eea

Substituting the matrix elements in Eq. (\ref{proton.pion}), one can analytically evaluate the OPE for
the correlator.  In $x$ representation for the correlator function we get:
\bea
\Pi_1(p_1^2,p_2^2) &=& \int d^4 x du e^{i p x} \left\{ e^{-i u q x} \times \right. 
\nnb \\
&-&\left\{ \frac{f_\pi}{\pi^4 x^6} \left[ (3 + 2 b + 3 b^2) \q + 4 b \frac{qx}{x^2} \x \right] \left(
\varphi_\pi(u) + x^2 g_1(u) \right) + \right.
\nnb \\
&+& \frac{f_\pi}{\pi^4 x^6} g_2(u)  (3 + 2 b + 3 b^2) i \left(\x - \frac{x^2}{qx}\q\right) -
\nnb \\
&-&  \frac{f_\pi}{288 \pi^2 x^4} \mu \qq \left( 16 (b-b^2) + \frac{1}{6} (-1 + 3 b - 2 b^2) m_0^2 x^2  
\right) \times 
\nnb \\
&\times& \left.  \varphi_\sigma(u) \left[ x^2 \q - qx \x \right] \right\} +
\nnb \\
&+& \int {\cal D} \alpha_i e^{-i (\alpha_1 + \bar u \alpha_3) qx} \left\{ \right.  
\nnb \\
&&i \frac{f_{3 \pi}}{24 \pi^2 x^2} \qq (b-b^2)(1-2u) \varphi_{3 \pi}(\alpha_i) 
\left[ 2 q x \q + q^2 \x \right] +
\nnb \\
&+& \frac{f_\pi}{2 \pi^4 x^6} b (1-2u) \varphi_\parallel(\alpha_i) \left[ -x^2 \q + 2 qx \x - q^2
\frac{x^2}{qx} \x \right] -
\nnb \\
&-& \frac{f_\pi}{2 \pi^4 x^6} \left[ \varphi_\perp(\alpha_i)(1-2u) - \tilde \varphi_\perp(\alpha_i) 
\right] b
\left( 2 q x + q^2 \frac{x^2}{qx} \right) \x + 
\nnb \\
&+&  \frac{f_\pi}{4 \pi^4 x^4} (b^2 + 1) \left( \q - \frac{qx}{x^2} \x \right) \tilde
\varphi_\parallel(\alpha_i) + 
\nnb \\
&+&\left. \left. \frac{f_\pi}{2 \pi^4 x^4} \frac{q^2}{qx} \tilde \varphi_\parallel(\alpha_i) b \x
\right\} \right\} +
\nnb \\
&+& \sigma_{\alpha \beta} \gamma_5 q^\alpha  \int d^4 x du e^{i px} x^\beta\left\{ e^{-i u q x} \right. 
\nnb \\
&&\left\{ \frac{f_\pi}{48 \pi^2 x^4} \qq \left( 16 (b-b^2) + \frac{1}{6} (-1 + 3 b - 2 b^2) m_0^2 x^2 \right) \times \right. 
\nnb \\
&\times& \left[ \varphi_\pi(u) + x^2 g_1(u) \right]  +
\nnb \\
&+& i \frac{f_\pi}{3 \pi^2 x^2} (b-b^2) \qq \frac{g_2(u)}{qx} - \left. \frac{f_\pi}{2 \pi^4 x^6} \mu (b^2-1) \varphi_\sigma(u)  \right\} +
\nnb \\
&+& \int {\cal D} \alpha_i e^{-i (\alpha_1 + \bar u \alpha_3) qx} \left\{ 
-\frac{f_\pi}{24 \pi^2 x^2} (b-b^2) \qq (1-2u) \varphi_\parallel(\alpha_i) \right. -
\nnb \\
&-& \left. \left. \frac{f_\pi}{24 \pi^2 x^2} \qq (b-1)  \tilde \varphi_\parallel (\alpha_i) \right\}
\right\} \label{p1}
\\ 
\Pi_2(p_1^2,p_2^2) &=& \int d^4 x du e^{i p x} \left\{ e^{-i u q x} \times \right. 
\nnb \\
&&\left\{- \frac{f_\pi}{2\pi^4 x^6} \left[ (b+1)^2 \q -2 (1+6b+b^2) \frac{qx}{x^2} \x \right] \left(
\varphi_\pi(u) + x^2 g_1(u) \right) + \right.
\nnb \\
&+& \frac{f_\pi}{2 \pi^4 x^6} g_2(u) (b+1)^2  i \left(\x - \frac{x^2}{qx}\q\right) +
\nnb \\
&+& \left. \frac{f_\pi}{576 \pi^2 x^4} \mu \qq \left( 16 + \frac{7}{6} m_0^2 x^2 
\right) (b^2-1) \varphi_\sigma(u) \left[ x^2 \q - qx \x \right] \right\} +
\nnb \\
&+& \int {\cal D} \alpha_i e^{-i (\alpha_1 + \bar u \alpha_3) qx} \left\{ \right.  
\nnb \\
&&-i \frac{f_{3 \pi}}{48 \pi^2 x^2} \qq (b^2-1)(1-2u) \varphi_{3 \pi}(\alpha_i) 
\left[ 2 q x \q + q^2 \x \right] -
\nnb \\
&-& \frac{f_\pi}{8 \pi^4 x^6} (1+6b+b^2) (1-2u) \varphi_\parallel(\alpha_i) \left[ -x^2 \q + 2 qx \x -
q^2
\frac{x^2}{qx} \x \right] +
\nnb \\
&+& \frac{f_\pi}{8 \pi^4 x^6} (1+6 b+b^2)\left[ \varphi_\perp(\alpha_i)(1-2u) - \tilde
\varphi_\perp(\alpha_i) \right] \left( 2 q x + q^2 \frac{x^2}{qx} \right) \x - 
\nnb \\
&-&  \frac{f_\pi}{8 \pi^4 x^4} \left[ (3 + 2 b + 3 b^2) \q - 4 (b+1)^2 \frac{qx}{x^2} \x \right] \tilde
\varphi_\parallel(\alpha_i) -
\nnb \\
&-& \left. \left. \frac{f_\pi}{8 \pi^2 x^4} \frac{q^2}{qx} \tilde \varphi_\parallel(\alpha_i) (1+6b +
b^2) \x
\right\} \right\} +
\nnb \\
&+& \sigma_{\alpha \beta} \gamma_5 q^\alpha  \int d^4 x du e^{i px} x^\beta\left\{ e^{-i u q x} \right. 
\nnb \\
&&\left\{- \frac{f_\pi}{96 \pi^2 x^4} \qq (b^2-1)\left( 16 + \frac{7}{6} m_0^2 x^2 \right) \left[ \varphi_\pi(u) +
x^2 g_1(u) \right] \right. -
\nnb \\
&-& i \frac{f_\pi}{6 \pi^2 x^2} (b^2-1) \qq \frac{g_2(u)}{qx} -
\left. \frac{f_\pi}{12 \pi^4 x^6} \mu (b-1)^2 \varphi_\sigma(u) \right\} +
\nnb \\
&+& \int {\cal D} \alpha_i e^{-i (\alpha_1 + \bar u \alpha_3) qx} \left\{ 
\frac{f_\pi}{48 \pi^2 x^2} (b^2-1) \qq (1-2u) \varphi_\parallel(\alpha_i) \right. -
\nnb \\
&-& \frac{f_\pi}{48 \pi^2 x^2} \qq (b^2-1)  \tilde \varphi_\parallel (\alpha_i) +
\nnb \\
&+& \left. \left. i \frac{f_{3\pi}}{4 \pi^4 x^6}(b-1)^2 (1-2u)qx \varphi_{3 \pi} (\alpha_i) \right\}
\right\}  \label{p2}
\eea
where in Eqs. (\ref{p1})-(\ref{p2}) we have neglected terms containing the gluon condensate $\gGgG$ as it gives a negligible 
contribution to the sum rules.

Our next problem is to perform Fourier transformation from $x$ to momentum representations and then, to get the result for
the theoretical part of the sum rules,  apply on the obtained result double Borel transformations on the variables
$p_1^2$ and $p_2^2$ in order to suppress the contributions of higher states and the continuum.
(the details can be found in [16-19]
). In calculating the Fourier transforms of terms containing
factors $qx$ or $1/qx$, we have performed integration by parts:
\bea
\int_0^1 du qx f(u) e^{-iuqx} &=& -i \int_0^1 du f'(u) e^{-i u qx} + f(u) e^{-i u qx} \ve_0^1  \\
\int_0^1 du \frac{e^{-i u qx}}{qx} g_2(u) &=& - i \int_0^1 du e^{-i u q x} G_2(u) - \frac{G_2(u)}{qx}
e^{-i u q x} \ve_0^1  \label{g2} \, ,
\eea
where $f'(u) = \frac{df}{dx}(u)$ and
\bea
G_2(u) = - \int_0^u du' g_2(u') \, .
\eea
The second term in Eq. (\ref{g2}) vanishes since $G_2(0) = G_2(1) = 0$.

After performing double Borel transformation over the variable $p_1^2=p^2$ and $p_2^2=(p+q)^2$ for the
$i \gamma_5 \not\!q$ structure, we obtain the following results for the theoretical part of the
sum rules:
\bea
\Pi_1^\gamma(M^2) &=&
-\frac{f_\pi}{8 \pi^2} M^6 f_2(x) [ ( 3 + 2 b + 3 b^2) \phi_\pi(u_0) - \frac{2 b}{3} u_0 \phi'_\pi
(u_0) ] +
\nnb \\
&+& (b^2 + 1) u_0 \phi'_\pi(u_0)] +
\nnb \\
&+& \frac{f_\pi}{\pi^2} M^4 f_1(x) (3 + 2 b + 3 b^2) \left[g_1(u_0) + G_2(u_0)\right] -
\nnb \\
&-& \frac{f_\pi}{4  \pi^2} u_0 g_2(u_0) M^4 f_1(x) (3 + 2 b + 3 b^2) g_2(u_0)  -
\nnb \\
&-& \frac{2 f_\pi \mu}{9} \qq (b-b^2) M^2 f_0(u_0) \left[ \phi_\sigma(u_0) + \frac{u_0}{2}
\phi'_\sigma(u_0) \right] +
\nnb \\
&+& \frac{f_\pi \mu}{216} m_0^2 \qq (-1 + 3 b - b^2) u_0 \phi'_\sigma(u_0) -
\nnb \\
&-& \frac{f_\pi}{\pi^2} M^4 f_1(x) u_0 g'_1(u_0) b -
\nnb \\
&-& \frac{f_\pi}{4 \pi^2} M^4 f_1(x) \left[u_0 I_1(1-2u,\phi_\parallel) + 2
I_{11}(1-2u,\phi_\parallel) \right] b -
\nnb \\
&-& \frac{f_{3\pi}}{3} \qq M^2 f_0(x) I_1(1-2u,\phi_{3\pi})  (b -
b^2) +
\nnb \\
&+& \frac{f_\pi}{4 \pi^2} M^4 f_1(x) u_0 \left[ I_1(1-2u,\phi_\perp) + I_1(1,\tilde \phi_\perp \right] b
-
\nnb \\
&-& \frac{f_\pi}{16 \pi^2} M^4 f_1(x) \left[ (b+1)^2 u_0 I_1(1,\tilde \phi_\parallel)  + 4 (b^2 +
1) I_{11}(1,\tilde \phi_\parallel) \right] +
\nnb \\
&+& \frac{f_\pi}{4 \pi^2} M^4 f_1(x) u_0 I_1(1,\tilde \phi_\perp) b \label{32}
\\
\Pi_2^\gamma(M^2) &=&
-\frac{f_\pi}{48  \pi^2} M^6 f_2(x) \left[ 3 (1+b)^2 \phi_\pi(u_0) + (1 + 6 b +
b^2) u_0 \phi'_\pi(u_0) \right] +
\nnb \\
&+& \frac{f_\pi}{2 \pi^2} M^4 f_1(x) \left[ g_1(u_0) + G_2(u_0) \right] (b+1)^2 -
\nnb \\
&-& \frac{f_\pi}{8 \pi^2}  M^4 f_1(x) u_0 g_2(u_0) (b+1)^2 +
\nnb \\
&+& \frac{f_\pi}{9} \mu \qq M^2 f_0(x)(b^2-1) \left[ \phi_\sigma(u_0) + \frac{u_0}{2} \phi'_\sigma
(u_0) \right] -
\nnb \\
&-& \frac{7 f_\pi \mu}{432} m_0^2 \qq u_0 \phi'_\sigma(u_0) (b^2-1) +
\nnb \\
&+& \frac{f_\pi}{4 \pi^2} M^4 f_1(x) u_0 g'_1(u_0) (1 + 6 b + b^2) +
\nnb \\
&+& \frac{f_\pi}{16 \pi^2} M^4 f_1(x) ( 1 + 6 b + b^2) \left[ u_0 I_1(1-2u,\phi_\parallel) + 
2 I_{11}(1-2u,\phi_\parallel) \right] +
\nnb \\
&+& \frac{f_{3 \pi}}{6} \qq M^2 f_0(x) (b^2-1) I_1(1-2u,\phi_{3 \pi}) -
\nnb \\
&-& \frac{f_\pi}{16 \pi^2}M^4 f_1(x) (1 + 6 b + b^2) I_1(1-2u,\phi_\perp) +
\nnb \\
&+& \frac{f_\pi}{8 \pi^2} M^4 f_1(x) \left[ (1+b)^2 u_0 I_1(1,\tilde \phi_\parallel) +  (3 + 2 b + 3
b^2) I_{11}(1,\tilde \phi_\parallel) \right] -
\nnb \\
&-& \frac{f_\pi}{16 \pi^2} M^4 f_1(x) (1 + 6 b + b^2) u_0I_1(1,\tilde \phi_\perp)
\, .
\eea
Similarly for the $\sig$ structure we get:
\bea
\Pi_1^\sigma(M^2) &=&
\frac{2 f_\pi}{3  \pi^2} \qq \left[ M^2 f_0(x)(b-b^2) -\frac{m_0^2}{24} (-1 + 3 b - b^2) \right]  \phi_\pi(u_0) -
\nnb \\
&-& \frac{8 f_\pi}{3} \qq (b - b^2) \left[ g_1(u_0) + G_2(u_0) \right]+
\nnb \\
&+& \frac{f_\pi}{8 \pi^2} \mu M^4 f_1(x) (b^2 - 1) \phi_\sigma(u_0) -
\nnb \\
&-& \frac{f_\pi}{3} \qq (b-b^2) I_{11}(1-2u,\phi_\parallel)  +
\nnb \\
&+& \frac{f_\pi}{3} \qq (b-1) I_{11}(1,\tilde \phi_\parallel)
\\
\Pi_2^\sigma(M^2) &=&
- \frac{f_\pi}{3} \qq \left[ M^2 f_0(x) -\frac{7 m_0^2}{24} \right] (b^2 - 1) \phi_\pi(u_0) +
\nnb \\
&+& \frac{4 f_\pi}{3} \qq (b^2 - 1) \left[ g_1(u_0) + G_2(u_0) \right] +
\nnb \\
&+& \frac{f_\pi}{48 \pi^2} \mu M^4 f_1(x) (b-1)^2 \phi_\sigma(u_0)+
\nnb \\
&+& \frac{f_\pi }{6} \qq (b^2-1) I_{11}(1-2u,\phi_\parallel) - 
\nnb \\
&+& \frac{f_{3 \pi}}{16 \pi^2} M^4 f_1(x) (b-1)^2 I_{11}(1-2u,\phi_{3 \pi})+
\nnb \\
&+& \frac{f_\pi}{6} \qq (b^2-1) I_{11}(1,\tilde \phi_\parallel) \label{35}
\eea
where the superscripts $\gamma$ and $\sigma$ corresponds to the $i \gamma_5 \not\!q$ and
$\sig$ structures respectively, $\mu = \frac{m_\pi^2}{2 m_q}
= -\frac{\la \bar q q \ra}{f_\pi^2}$,
$x=\frac{s_0}{M^2}$, $f_n(x) = 1 - e^{-x} (1 + x + \frac{x^2}{2} + \ldots + \frac{x^n}{n!})$,
$s_0$ is the continuum threshold, $u_0 = M_2^2/(M_1^2 + M_2^2)$,
\bea
\frac{1}{M^2} = \frac{1}{M_1^2} + \frac{1}{M_2^2}
\eea
and the function $I_1$ and $I_{11}$ are defined as:
\bea
I_{11}(f(u),g(\alpha_i)) &=& \int_0^{u_0} d \alpha_1 \int_0^{1-u_0} d \alpha_2 \frac{f(\frac{u_0 -
\alpha_1}{1-\alpha_1 - \alpha_2})}{1-\alpha_1-\alpha_2} g(\alpha_1,\alpha_2,1-\alpha_1-\alpha_2) 
\nnb \\ \\
I_1(f(u),g(\alpha_i)) &=& \int_0^{u_0} d \alpha_1 \int_0^{1-u_0} d \alpha_2 \frac{f(\frac{u_0 -
\alpha_1}{1-\alpha_1 - \alpha_2})}{1-\alpha_1-\alpha_2} g(\alpha_1,\alpha_2,1-\alpha_1-\alpha_2) +
\nnb \\
&+& \int_0^{1-u_0} d \alpha_2 \frac{f(0)}{1-u_0-\alpha_2} g(u_0,\alpha_2,1-u_0-\alpha_2) -
\nnb \\
&-& \int_0^{u_0} d \alpha_1 \frac{f(1)}{u_0-\alpha_1} g(\alpha_1, 1-u_0,u_0-\alpha_1)
\eea

Equating theoretical (see Eqs. (\ref{32}-\ref{35}))and phenomenological (see Eq. (\ref{phenomenological}))  parts of the
correlator function (\ref{correlation}) we
arrive at the following sum rules for the meson baryon coupling constants: \\
\begin{itemize}
\item[a)] For the $i \gamma_5 \not\!q$ structure:
\bea
-m_N \lambda_N^2 g_{\pi N N} e^{-\frac{m_N^2}{M^2}} &=& \Pi_1^\gamma(M^2) +\Pi_2^\gamma(M^2) \nnb \\
-m_\Sigma \lambda_\Sigma^2 g_{\pi \Sigma \Sigma} e^{-\frac{m_\Sigma^2}{M^2}} &=&
\Pi_1^\gamma(M^2 ) \nnb \\
-m_\Xi \lambda_\Xi^2 g_{\pi \Xi \Xi} e^{-\frac{m_\Xi^2}{M^2}} &=& - \Pi_2^\gamma(M^2) \nnb \\
-\sqrt{3} m_N \lambda_N^2 g_{\eta N N} e^{-\frac{m_N^2}{M^2}} &=& \Pi_1^\gamma(M^2) - 
\Pi_2^\gamma(M^2) \nnb \\
-\sqrt{3} m_\Sigma \lambda_\Sigma^2 g_{\eta \Sigma} e^{-\frac{m_\Sigma^2}{M^2}} &=& \Pi_1^\gamma(M^2) +
2 \Pi_2^\gamma(M^2) \nnb \\
-\sqrt{3} m_\Xi \lambda_\Xi^2 g_{\eta \Xi \Xi} e^{-\frac{m_\Xi^2}{M^2}} &=& -2 \Pi_1^\gamma(M^2) -
\Pi_2^\gamma(M^2) \label{srgamma} 
\eea

\item[b)] For the $\sig$ structure:
\bea
\lambda_N^2 g_{\pi N} e^{-\frac{m_N^2}{M^2}} &=& \Pi_1^\sigma(M^2) + \Pi_2^\sigma(M^2) \nnb \\
\lambda_\Sigma^2 g_{\pi \Sigma \Sigma} e^{-\frac{m_\Sigma^2}{M^2}} &=& \Pi_1^\sigma(M^2) \nnb \\
\lambda_\Xi^2 g_{\pi \Xi \Xi} e^{-\frac{m_\Xi^2}{M^2}} &=& - \Pi_2^\sigma(M^2) \nnb \\
\sqrt{3} \lambda_N^2 g_{\eta N} e^{-\frac{m_N^2}{M^2}} &=& \Pi_1^\sigma(M^2) - \Pi_2^\sigma(M^2) \nnb \\
\sqrt{3} \lambda_\Sigma^2 g_{\eta \Sigma \Sigma} e^{-\frac{m_\Sigma^2}{M^2}} &=&
\Pi_1^\sigma(M^2) + 2 \Pi_2^\sigma(M^2) \nnb \\
\sqrt{3} \lambda_\Xi^2 g_{\eta \Xi \Xi} e^{-\frac{m_\Xi^2}{M^2}} &=& -2 \Pi_1^\sigma(M^2) -
\Pi_2^\sigma(M^2) \label{srsigma}
\eea
\end{itemize}

These results are obtained  in the $SU(3)$ limit,i.e. all masses and residues of the baryons are the
same 
and $\ss = \la \bar q q \ra$ ($q=u,d$).  Note also that Eqs. (\ref{srgamma}) and (\ref{srsigma}) are
consistent with the $SU(3)$ symmetry relations Eq. (\ref{su3}).  In terms of the OPE, the F/D ratio can
be identified as:
\bea
2 \alpha = \frac{\Pi_1(M^2)}{\Pi_1(M^2) + \Pi_2(M^2)} \rightarrow
F/D = \frac{\Pi_1(M^2)}{\Pi_1(M^2) +2 \Pi_2(M^2)}
\eea

\section{Numerical Analysis}
In this section we analyze the sum rules obtained in the previous section for the coupling constants in
the $SU(3)$ limit and study the dependence of the $F/D$ ratio on the Dirac structure.  Since the
coupling constants are  physical quantities, they should be independent of the parameter $b$ and the
continuum threshold $s_0$.  Therefore our first problem is to find the region in the parameter
space where they are practically independent of $b$ and $s_0$.  In $SU(3)$ limit, all baryon masses and
their residues are equal and also $ \la q \bar q \ra=\ss$ and $f_\eta=f_\pi$.  

The main input parameters of the sum rules (see Eqs. (33-36) and (40-41)), are the pion wave functions.
In \cite{chernyak2}, a theoretical framework has been developed to study these functions.  The leading twist 2 pion wave function
can be expressed as an expansion in Gegenbauer polynomials $C_i^{3/2}$ \cite{braun0}:
\bea
\varphi_\pi (u) = 6 u (1-u) \left[ 1 + a_2 C_2^{3/2}(2u-1) + a_4 C_4^{3/2} + \ldots \right] \label{expansion}
\eea
The coefficients $a_i$ renormalize multiplicatively. On the basis of the approximate conformal symmetry of QCD,
it has been shown in \cite{braun05} that the expansion (\ref{expansion}) converges sufficiently fast so that terms 
with $n > 4$ are negligible.

In the calculations  we have used the following forms of the
wavefunctions appearing in the meson matrix elements (see e.g. \cite{braun0, braun05} for more details):
\bea
\varphi_\sigma (u,\mu)&=&6u\bar{u}\Big[ 1+C_2 \frac{3}{2}(5(2 u- 1)^2-1) \nnb \\ 
&&{}+C_4 \frac{15}{8}(21(2 u - 1)^4-14(2 u-1)^2+1)\Big] ~, \nnb \\
\varphi_\pi(u) &=& 6 u (1-u) \left[ 1 + a_2 \frac{3}{2} \left( 5 (2 u -1)^2 - 1 \right) + 
a_4 \frac{15}{8} \left( 21 (2 u-1)^4 - 14 (2 u -1) +1 \right) \right] \nnb \\
\varphi_P(u)&=& 1+B_2\frac{1}{2}(3(2 u-1)^2-1) +B_4\frac{1}{8} (35(2 u-1)^4 \nnb\\ &&
{}-30(2 u-1)^2+3) ~, \nnb \\
g_1(u)&=&\frac{5}2\delta^2 (1-u)^2u^2+\frac{1}{2}\varepsilon
\delta^2 [(1-u) u(2+13(1-u)u)+10 u^3\ln u(2-3u+\frac{6}{5} u^2) \nnb \\ 
&&{}+10 (1-u)^3\ln (1-u)(2-3(1-u)+\frac{6}{5} (1-u)^2)] ~, \nnb \\
g_2(u) &=& \frac{10}{3} \delta^2 u (1-u) (2 u -1) \nnb \\
\varphi_\parallel (\alpha_i) &=& 120\delta^2 \varepsilon (\alpha_1-\alpha_2)\alpha_1\alpha_2\alpha_3 \nnb \\
\varphi_\perp (\alpha_i) &=& 10 \delta^2 (\alpha_1 - \alpha_2) \alpha_3^2 \left[ 1 + 6 \varepsilon (1-2\alpha_3) \right] \nnb \\
\tilde\varphi_\parallel (\alpha_i) &=& - 40 \delta^2 \alpha_1 \alpha_2 \alpha_3 \left[ 1 + 3 \varepsilon (1-3\alpha_3) \right] \nnb \\
\tilde\varphi_\perp(\alpha_i) &=& 10 \delta^2 (1-\alpha_3) \alpha_3^2 \left[1 + 6 \varepsilon (1-2\alpha_3) \right] \nnb \\
\varphi_{3 \pi} (\alpha_i) &=& 360 \alpha_1 \alpha_2 \alpha_3^2
\eea
where 
$\delta$ is defined by matrix element:
\bea
\la \pi \ve g_s \bar q \tilde G_{\alpha \mu} \gamma^\alpha q \ve 0 \ra = i \delta^2 f_\pi q_\mu \, .
\eea
and 
\begin{eqnarray}
B_2=30\frac{f_{3\pi}}{\mu f_\pi}\,, ~~
B_4=\frac{3}{2}\frac{f_{3\pi}}{\mu f_\pi} 
(4\omega_{2,0}-\omega_{1,1}-2\omega_{1,0})\,,
\nonumber\\
C_2=\frac{f_{3\pi}}{\mu f_\pi}(5-\frac{1}{2} \omega_{1,0})\,,~~
C_4=\frac1{10}\frac{f_{3\pi}}{\mu f_\pi}(4\omega_{2,0}-\omega_{1,1})~.
\end{eqnarray}
The additional parameters appearing in the above are numerically given by
$$ 
\omega_{1,0} = -2.88\,
,~~ \omega_{2,0}= 10.5\,,~~
\omega_{1,1} = 0\,,~~
\varepsilon =0.5 \, ,~~ a_2 = \frac{2}{3} \, ,~~ a_4 = 0.43 \, , 
$$ 
which corresponds to choosing the renormalization scale at $1 ~ GeV$.

In our calculations, we set $M_1^2=M_2^2=2 M^2$ since the initial and final baryons are identical, 
which corresponds to setting $u_0=1/2$.
Hence, in the sum rules, only the value of the wavefunctions at the symmetry point $u=1/2$ are needed.

The values of the other input parameters appearing in the sum rules are: 
$f_\pi =0.013 \, GeV$, $\la g^2 G^2 \ra = 0.474 \, GeV^4$, $\la \bar q q \ra = - (0.243)^3 \,
GeV^3$, $m_0^2 = (0.8 \pm 0.2)\, GeV^2$ \cite{belyaev3}, $\delta^2=0.2 \, GeV^2$ \cite{novikov}, $f_{3
\pi} = 0.0035
\, GeV^2$, $\mu = 1.8797$. 

In Figs. (1) and (2), we present the dependence of $g_{\pi NN} \lambda^2_N(b)$ on the
Borel mass $M^2$ at three different values of $b$, namely $b=-1.5$, $b=1.5$ as well as $b=-1$ which corresponds to the
Ioffe current for the structures $i \gamma_5 \not\!q$ and $\sig$ respectively.  The continuum
threshold is chosen to be $s_0=2.07 \, GeV^2$ corresponding to the Roper resonance.  To analyze the sensitivity of the sum
rules to the continuum threshold, we also plotted the results for the value $s_0 = 2.57 \, GeV^2$.  
From these figures we obtained that the working region for
the Borel mass is $0.6 \le M^2 \le 1.2 \, , GeV^2$.  
Also we see that the tensor structure 
$\sig$ is more stable with respect to variations of the Borel mass
and also the variations of the continuum threshold for all curves.  For example at $M^2=1 \, GeV^2$, the
results are practically independent of $s_0$ and $b$.  The results change about $\sim 5\%$ for the
structure $\sig$ with variations of $s_0$ and $b$. This indicates
that the results obtained from the structure $i \gamma_5 \not\!q $ are less reliable.  

In Figs (3) and (4), the dependence of $g_{{\cal M}BB} \lambda^2_B(b)$ on the
parameter $b$ at $M^2=1 \, GeV^2$ and at $s_0=2.07 \, GeV^2$ is presented for the structures $i \gamma_5
\not\!q$ and $\sig$.  The coupling constants
$g_{{\cal M}BB}$ are physical parameters and hence it should not depend on the arbitrary parameter
$b$.  Since $\lambda_B$'s are the same for all baryons due to the $SU(3)$ symmetry, one expects the
graphs to be just multiples of one another.  The results obtained from the
structure $i \gamma_5 \not\!q $ are far from satisfying this criteria.  On the other hand the
results obtained form the $\sig$ structure are all zero at $b=1$
similar to the traditional sum rules(see \cite{doi}) but the second zero position is different for 
different coupling constants. Note that the region $-0.5 \le b \le 1$ is unphysical since in this region
the mass sum rules yields a negative value for $\lambda_B^2(b)$ \cite{doi}.
As we see from Fig. 4, for each sum rule, this unphysical region becomes wider and contains the region
between the two zeroes, since the sign of $g_{{\cal M} BB} \lambda_B^2(b)$ should be the same as the
sign of $g_{{\cal M} BB}$, hence it should not change. From these figures also
one is led to the conclusion that the predictions of the $i \gamma_5 \not\!q $ structure is not as
reliable as the predictions of the $\sig$ structure in constructing sum rules. 

In Fig (5), the dependence of the F/D ratio for the above mentioned structures on  $cos\theta$ is
presented.  Here, $\theta$ is defined as $b = \tan \theta$ and only the physical region for the
parameter $\theta$ is shown.  From these figures, it follows that for the tensor structure, the ratio
$F/D$ is practically independent on the continuum threshold, but for the $i \q$ structure, it has a
strong dependence on $s_0$.  For this reason, prediction for the $F/D$ ratio from the tensor structure
is more reliable. The dependence of the $F/D$ ration on the Borel parameter $M^2$ also turns out to be
very weak. Being a physical parameter, the F/D ratio should be independent of $\cos \theta$.  From the
figure, we see that F/D is quite stable (for the tensor structure) in the region $-0.25 \le \cos \theta
\le 0.50$. This region is also away from the unphysical region for $b$. In this region,
we obtain the result  $F/D=0.6 \pm 0.1$. Note that $SU(6)$ quark model predict $F/D=2/3$.  Analysis of
semileptonic decay of hyperons give $F/D \simeq 0.57$ \cite{ratcliffe} and the traditional sum rules
yield $0.6 \le F/D \le 0.8$ \cite{doi}.  Within errors our results are in a good agreement with 
all existing results.
\newpage

\newpage
\section*{Figure Captions}
\begin{itemize}
\item[Fig. 1.] The dependence of $g_{\pi N N} \lambda^2_N(b)$ on the Borel mass, $M^2$ at
$b=-1.5, \,- 1.0, \, 1.5$ and at the continuum threshold $s_0=2.07 \, GeV^2$ and $s_0=2.57\, GeV^2$ for the
$ i\gamma_5 \not\!q$ structure.
\item[Fig. 2.] The same as Fig. 1 but for the structure $\sig$
\item[Fig. 3.] The dependence of $g_{{\cal M} B B} \lambda^2_B(b)$ on $b$ at $M^2=1 \, GeV^2$ and the
continuum threshold $s_0=2.07 \, GeV^2$ for the $i\gamma_5\not\!q $ structure.
\item[Fig. 4.] The same as Fig. (3) but for the structure $\sig$ 
\item[Fig. 5.] The dependence of F/D on $\theta$ at $M^2=1 \, GeV^2$ for both structures at two  values of the
continuum threshold, $s_0= 2.07 \, GeV^2$ and $s_0=2.57 \, GeV^2$.
\end{itemize}

\newpage

\newpage
\begin{figure}
\vskip 8.0 cm
    \includegraphics{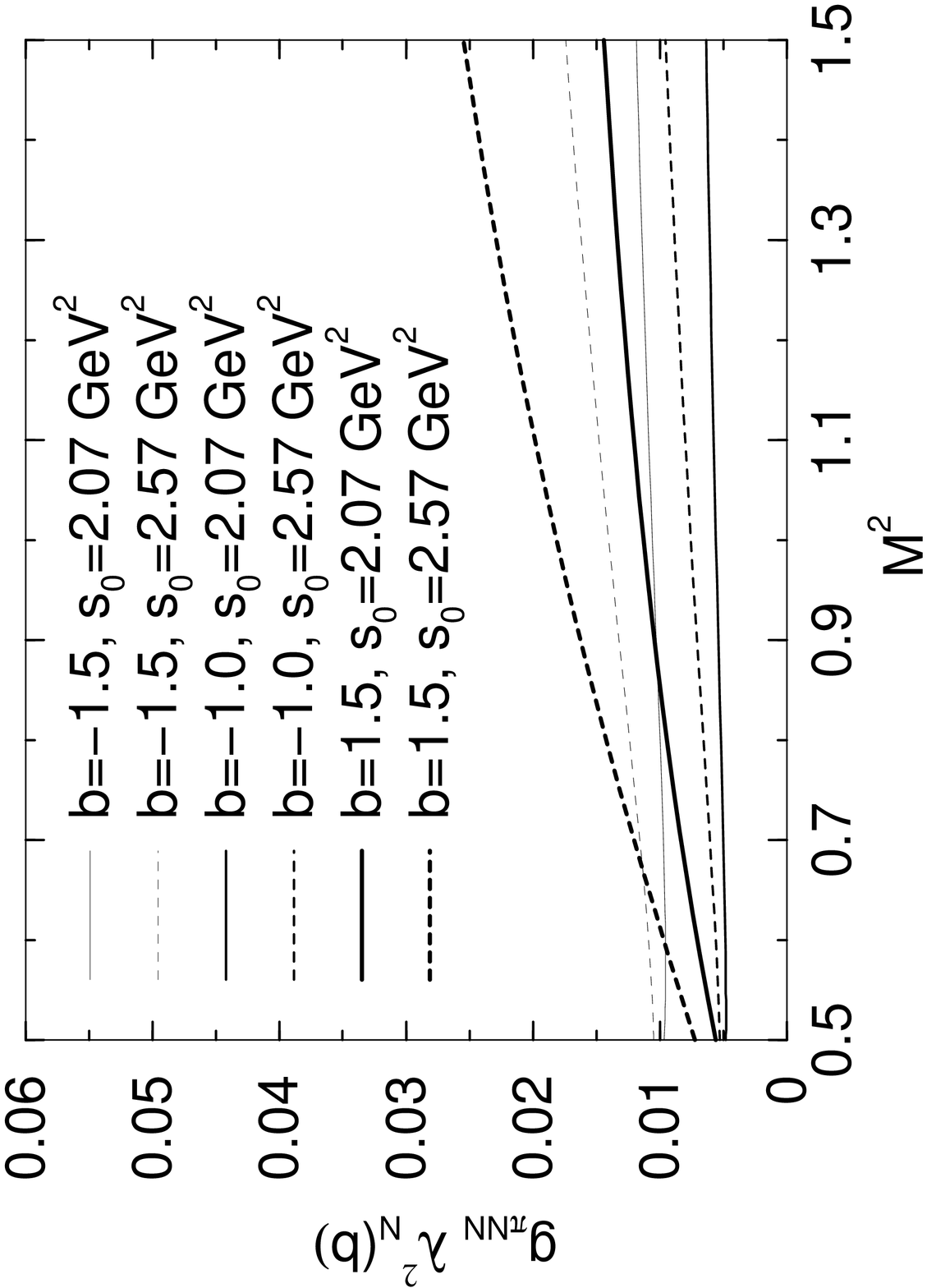}
\vskip -2.5cm
\caption{}
$\left. \right.$
\vspace{3cm}
\end{figure}
\begin{figure}
$\left. \right.$
\vskip 4.5 cm
    \includegraphics{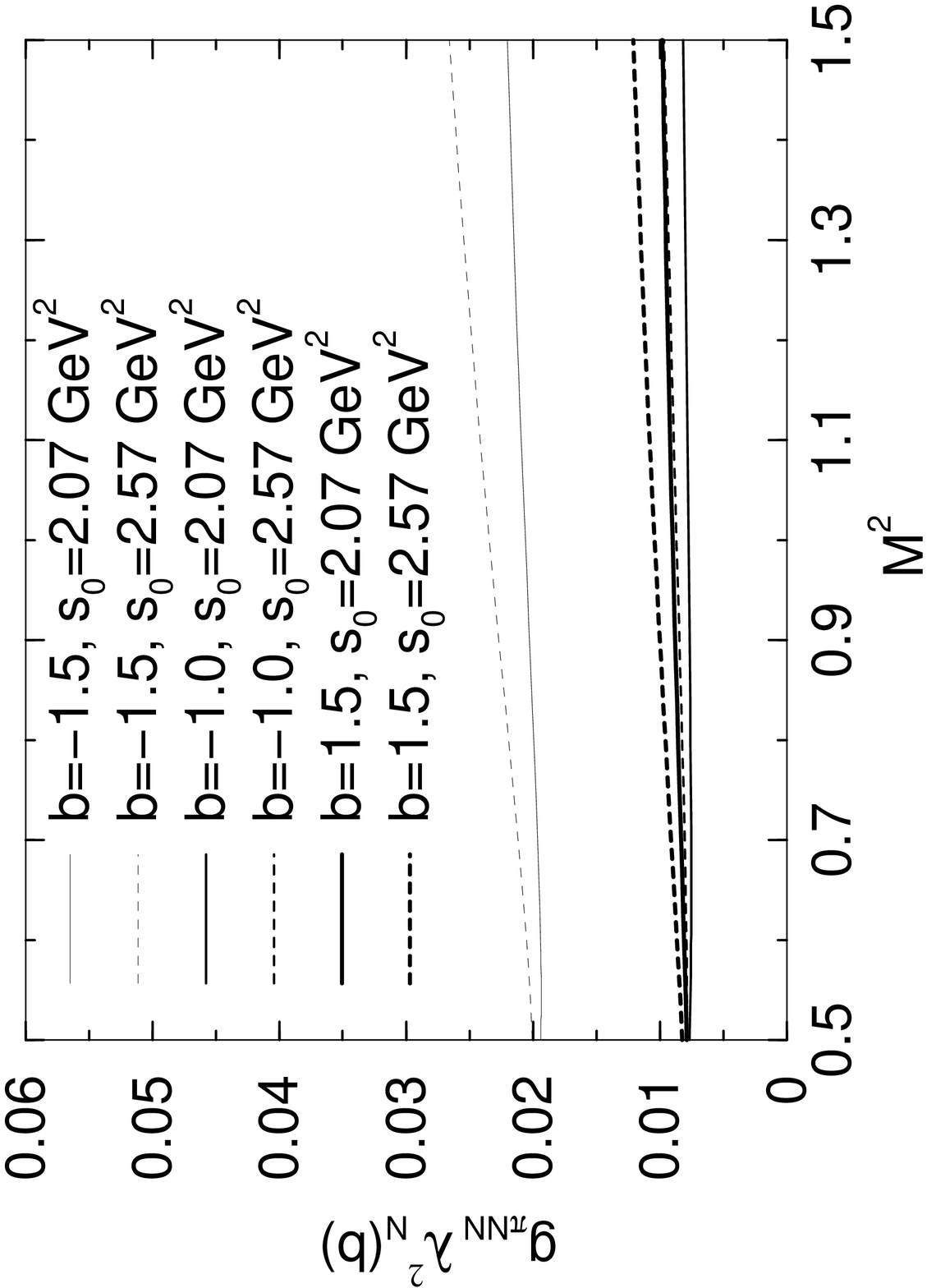}
\vskip -2.5cm
\caption{}
\end{figure}
\newpage
\begin{figure}
\vskip 7.5 cm
    \includegraphics{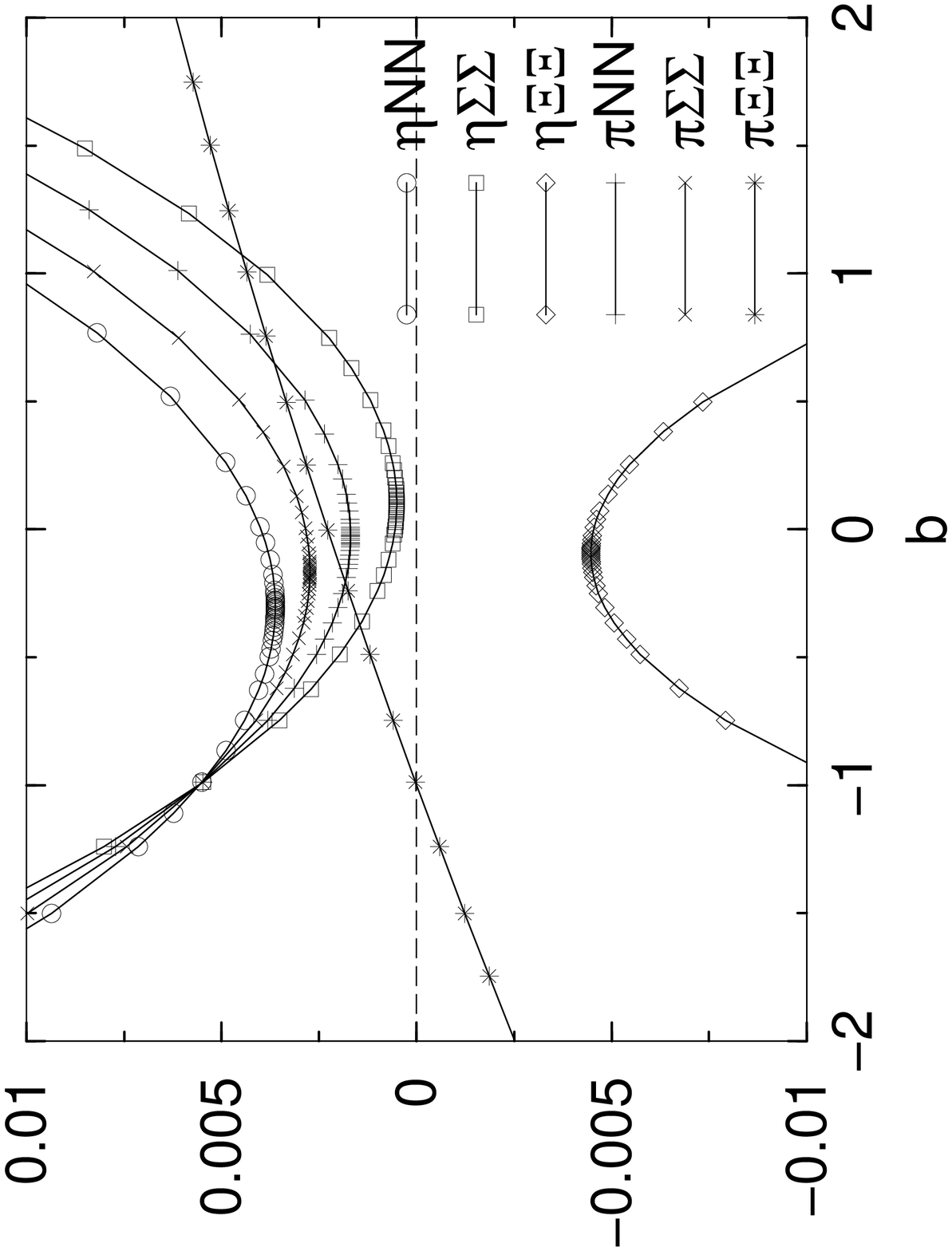}
\vskip -2.1cm
\caption{}
\vspace{2cm}
\end{figure}
\begin{figure}
\vskip 7.5 cm
    \includegraphics{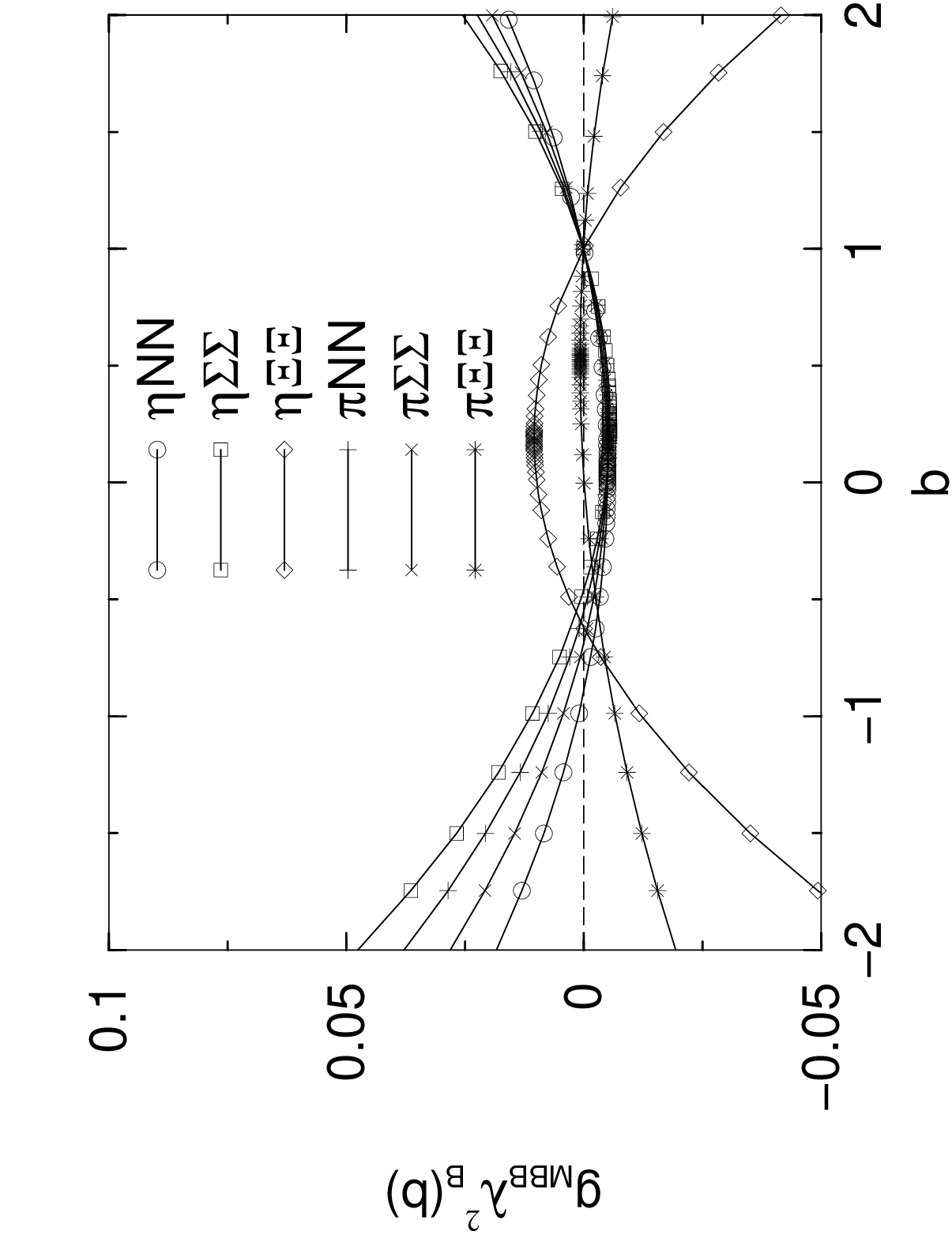}
\vskip -2.1cm
\caption{}
\end{figure}
\newpage
\begin{figure}
\vskip 7.5 cm
    \includegraphics{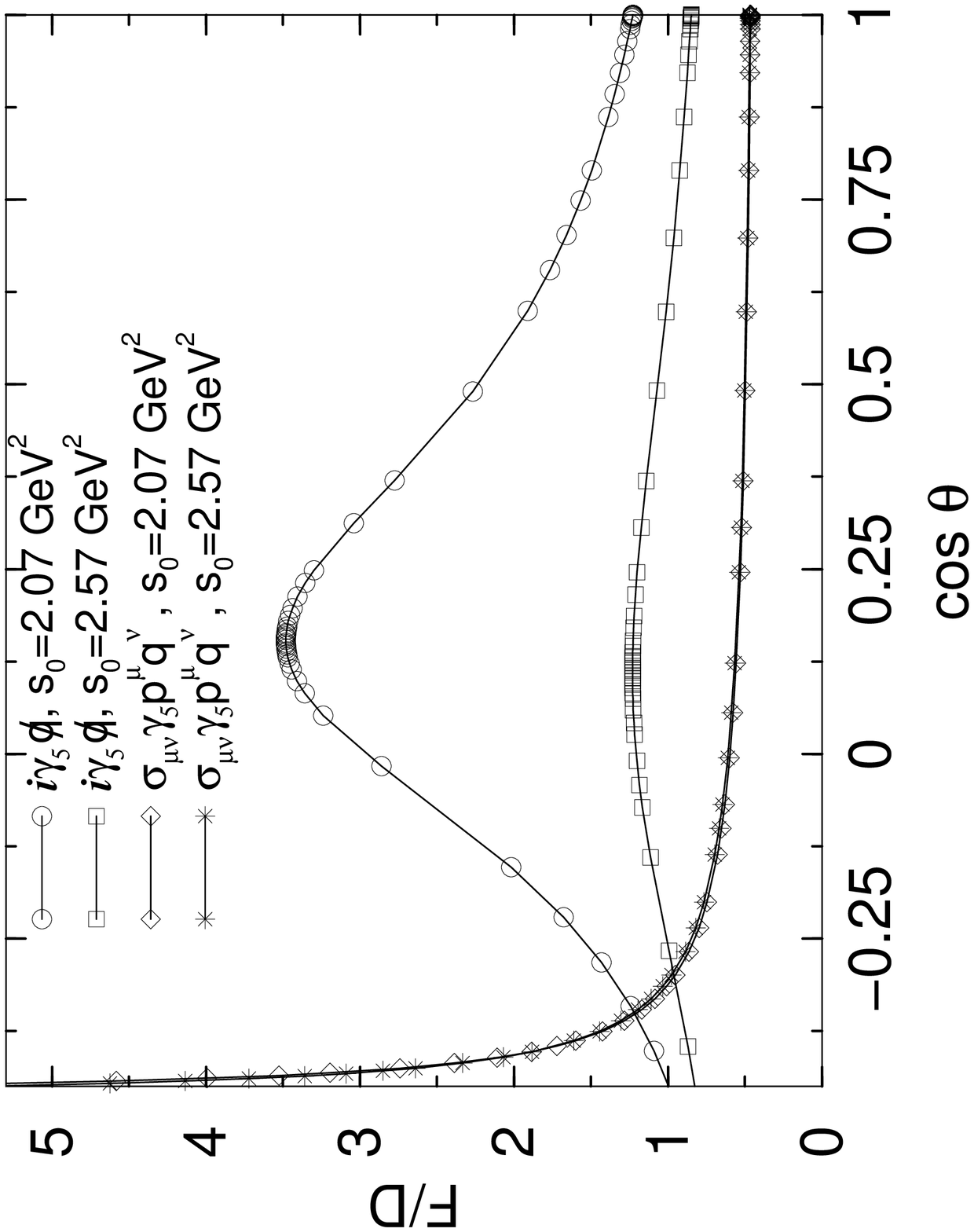}
\vskip 1.1cm
\caption{}
\vspace{2cm}
\end{figure}

\end{document}